\documentclass[11pt,a4paper]{article}
\pdfoutput=1

\usepackage{amsmath}
\usepackage{amssymb}
\usepackage{cite}
\usepackage{graphicx}
\usepackage{bbm}

\newcommand{\be}{\begin{equation}}  
\newcommand{\ee}{\end{equation}}  
\newcommand{\ol}[1]{\overline{#1}}
\newcommand{\hc}{+\,\mathrm{h.c.}}
\newcommand{\vev}[1]{\langle #1 \rangle}

\newcommand{\SU}[1]{\ensuremath{\mathrm{SU}(#1)}}
\newcommand{\U}[1]{\ensuremath{\mathrm{U}(#1)}}
\newcommand{\into}{\ensuremath{\,\rightarrow\,}}

\begin{document}

\thispagestyle{empty}

\begin{flushright}
  DESY 11-070 \\
  May 2011\\
\end{flushright}
\vspace*{2.5cm}

\begin{center}

{\LARGE\bf  Light Higgsinos as Heralds\\ of Higher-Dimensional
Unification\\[10mm]}

{\large F.~Br\"ummer and W.~Buchm\"uller}\\[5mm]

{\it
Deutsches Elektronen-Synchrotron DESY,\\ 
Notkestra\ss e 85, D-22607 Hamburg, Germany\\[3mm]
}

\vspace*{12mm}

\begin{abstract}
\noindent Grand-unified models with extra dimensions at the GUT scale will
typically contain exotic states with Standard Model charges and GUT-scale
masses. They can act as messengers for gauge-mediated supersymmetry breaking. If
the number of messengers is sizeable, soft terms for the visible sector fields
will be predominantly generated by gauge mediation, while gravity mediation can
induce a small $\mu$ parameter. We illustrate this hybrid mediation pattern
with two examples, in which the superpartner spectrum contains
light and near-degenerate higgsinos with masses below $200$ GeV. The typical 
masses of all other superpartners are much larger, from at least $500$ GeV up
to several TeV. The lightest superparticle is the gravitino, which may be the
dominant component of dark matter.

\end{abstract}

\end{center}

\clearpage

\section{Introduction}

Supersymmetry, extra space-time dimensions, and grand unification are among the
most promising proposals for physics beyond the Standard Model. These three
paradigms are elegantly incorporated in various modern approaches to unified
model building, such as heterotic string compactifications, F-theory models, or
purely field-theoretic orbifold GUTs. Extra dimensions opening up around the GUT
scale provide new mechanisms to break the GUT gauge symmetry and to solve the
problems of conventional four-dimensional SUSY GUTs, such as doublet-triplet
splitting or proton decay. They are necessary ingredients in unifying the
Standard Model forces with gravity within string theory.

The gauge couplings in the Minimal Supersymmetric Standard Model (MSSM) unify at
$M_{\rm GUT}\approx 10^{16}$ GeV, which sets the scale where the GUT gauge
symmetry should be broken and around which unwanted exotic states will decouple.
In conventional four-dimensional GUTs, all states fall into complete GUT
multiplets at scales above $M_{\rm GUT}$. This is in general not true in
higher-dimensional constructions. In orbifold GUTs, for instance, the massless
spectrum generally contains incomplete GUT multiplets. They can originate from
the untwisted or twisted sector of an underlying heterotic string model. Likewise, 
F-theory GUTs can give rise to split multiplets when breaking the unified gauge group
with hypercharge flux.

Obtaining a pair of Higgs doublets without their colour triplet partners is of
course very welcome. Any additional split multiplets, however, should pick up
masses not too far from $M_{\rm GUT}$ in realistic models, in order not to
affect gauge coupling unification.\footnote{An exception here might be F-theory
models, in which gauge couplings are not actually predicted to unify at $M_{\rm
GUT}$ \cite{Blumenhagen:2008aw}.} This can be achieved by Standard Model singlet
fields acquiring expectation values of the order $M_{\rm GUT}$, thus giving
masses to vector-like pairs of exotics. If these singlets also obtain $F$-term
expectation values from coupling to the SUSY breaking hidden sector, the
vector-like exotics will act as messenger fields for gauge-mediated SUSY
breaking. The resulting patterns of soft SUSY breaking terms will be rather
different from that of conventional low-scale gauge mediation, where the
messengers would need to form complete GUT multiplets. Here, with the messenger
scale close to the GUT scale, the presence of split messenger multiplets
generically leads to non-universal gaugino masses.\footnote{Similar observations
were made in the context of F-theory models in \cite{Marsano:2009wr}. More
recently, in \cite{Anandakrishnan:2011zn} messengers in split multiplets were
used in heterotic models to improve the precision of gauge coupling
unification.}

Moreover, with a messenger scale $M_{\rm GUT}$, gauge-mediated contributions to
soft SUSY breaking terms are comparable with gravity-mediated soft terms. This
is because the latter are $M_{\rm P}$-suppressed, while the gauge-mediated terms
are loop-suppressed, and the GUT scale is lower than the Planck scale by roughly
a loop factor. For a sizeable number of messengers, gauge-mediated terms will
dominate. Gauge mediation does not, however, give rise to a $\mu$ term (unless
specific Higgs-messenger couplings are introduced, which in turn would induce a
too large $B_\mu$). Instead $\mu$ can be generated, as in gravity mediation, by
the Giudice-Masiero mechanism \cite{Giudice:1988yz}. It will thus be of the
order of the gravitino mass, and smaller than the soft masses if these are
predominantly of gauge mediation origin. The same applies to $B_\mu$ and to the
trilinear $a$-terms, which are suppressed in gauge mediation. The GUT-scale MSSM
parameters are then characterised roughly by the hierarchy 
\be\begin{split}
\{\mu,\,m_{3/2},\,a,\,\sqrt{B_\mu}\}\,\ll\,\{M_{1/2},m_0,\,m_H\}\,.
\end{split}
\ee 
Here $a$ stands collectively for any trilinear $a$-parameters, $m_0$ for squark
and slepton soft masses, $m_H$ for Higgs soft masses, and $M_{1/2}$ for gaugino
masses. 

Current experimental limits from chargino searches are consistent with the
``gravity-mediated'' terms on the LHS around $100$ GeV, whereas the
``gauge-mediated'' terms on the RHS may be TeV or larger. This leads to a quite
peculiar low-energy spectrum with the following main features: The only light
states with masses around 100 GeV are Higgs and higgsino fields and the
gravitino. The gravitino can naturally be the lightest supersymmetric particle
(LSP), with a higgsino-like neutralino NLSP. The second neutralino and a
higgsino-like chargino are slightly heavier. The mass of the lightest Higgs
scalar is lifted to around $120$ GeV by large squark loop effects. All the
remaining states, that is the heavier Higgs bosons, squarks, sleptons,
gluinos, and gaugino-like chargino and neutralinos, have masses of at least
around $500$ GeV and up to several TeV (the only possible exception being one of
the scalar taus, which can be somewhat lighter).

The resulting LHC physics is quite rich and distinctive, and will be the subject
of a future study. Here we merely point out that, in our scenario, most of the
new MSSM particles are necessarily beyond the reach of early LHC searches. In
particular, all coloured states are predicted to be rather heavy. This is
consistent with the null result of the first LHC SUSY searches so far
\cite{Khachatryan:2011tk,Aad:2011hh}.

The stable gravitino serves as a natural dark matter candidate, allowing for a
high reheating temperature as required by thermal leptogenesis. With standard
cosmology and if R-parity is conserved, the higgsino NLSP is long-lived, and one
has to make sure that its late-time decay does not destroy the successful
predictions of primordial nucleosynthesis. Such a picture was first advocated in
\cite{Bolz:1998ek}, and has recently been studied for a general neutralino LSP
in \cite{Covi:2009bk}. In our case the NLSP abundance is significantly reduced
by coannihilation with charginos, since the lighter chargino is nearly
mass-degenerate with the NLSP. It turns out, however, that this effect alone is
not sufficient to evade the stringent BBN bounds. As we shall see, our model can
still be made consistent with early-universe cosmology with some mild
modifications, such as a small amount of R-parity violation, or some moderate
entropy production before nucleosynthesis.

\section{Soft terms from hybrid gauge-gravity mediation}\label{softterms}

The simplest models of gauge-mediated supersymmetry breaking (see
\cite{Giudice:1998bp} for a review) contain a background chiral superfield $X$
which breaks supersymmetry, as well as some massive messenger fields in
vector-like pairs $\Sigma_i$, $\widetilde\Sigma_i$. There is a superpotential
\be\label{mingmW}
W=\sum_i \lambda_i\,X\Sigma_i\widetilde\Sigma_i
\ee
which, after $X$ is set to its expectation value $\vev X=M_{\rm m}+F\theta^2$,
gives rise to both a supersymmetric messenger mass $M_{\rm m}$ and a
SUSY-breaking mass splitting. Provided that $F\ll M_{\rm m}^2$, the contribution
of each messenger pair to the gaugino masses at the messenger scale is 
\be
M_a=\frac{g_a^2}{16\pi^2}n_a(r_i)\,\frac{F}{M_{\rm m}}\,.
\ee
Here $a=1,\,2,\,3$ labels the Standard Model gauge factors, $g_a$ is the running
gauge coupling, and the messengers are in representations  $r_i$ and $\ol r_i$
with Dynkin index $n_a(r_i)$. 

In addition there is also a gravity-mediated contribution to the soft masses.
For instance, if the gauge kinetic functions depend on $X$ as in
\be
{\cal L}=\frac{1}{4}\sum_a \int d^2\theta\,\left(\frac{1}{g_a^2}+\kappa_a
\frac{X}{M_{\rm P}}\right)\,W^{a\alpha} W^a_{\alpha}\hc\,,
\ee
then there will be gravity-mediated terms
\be
M_a=\frac{1}{2} g_a^2\kappa_a\,\frac{F}{M_{\rm P}}\,,
\ee
with the Planck scale $M_{\rm P}=2.4\cdot 10^{18}$ GeV.\footnote{The
$X$-dependence also affects the gauge coupling through the vev for the lowest
component of $X$. We neglect this since we are interested in the case $M_{\rm
m}/M_{\rm P}\ll 1$.}

For couplings $\kappa_a$ of order one, the contribution of each messenger pair
is comparable with the gravity-mediated piece if the messenger masses are
roughly a loop factor below the Planck scale. For a large number of messengers
gauge mediation dominates.

Similar statements hold for soft scalar masses. The gauge-mediated soft mass for
a chiral supermultiplet $\Phi$ is
\be
m_\Phi^2=\frac{2}{\left(16\pi^2\right)^2}\left(\sum_{ai}
g_a^4\,C_a\,n_a(r_i)\right)\left|\frac{F}{M_{\rm m}}\right|^2\,,
\ee
where $C_a$ is the quadratic Casimir for the representation under which $\Phi$
transforms. Gravity-mediated pieces are induced by operators 
\be
{\cal L}=\int d^4\theta\,\left(\frac{X^\dag}{M_{\rm
P}}\hc-\frac{1}{2}\frac{X^\dag X}{M_{\rm P}^2}\right)\,\Phi^\dag\Phi\,,
\ee
with ${\cal O}(1)$ couplings omitted, which gives
\be
m_\Phi^2=\frac{1}{2}\left|\frac{F}{M_{\rm P}}\right|^2\,.
\ee
As for the gaugino masses, the typical mass scale for the contribution of each
messenger pair is $F/(16\pi^2 M_{\rm m})$, while the typical mass scale for the
gravity-mediated piece is $F/M_{\rm P}$. They are comparable if $M_{\rm
m}\approx M_{\rm P}/(16\pi^2)$. Gauge mediation dominates if the number of
messengers is sizeable. Note that scalar masses grow roughly with the square
root of the messenger number; this is unlike the gaugino masses, which grow
linearly with the messenger number and hence tend to be larger if the latter is
large.

The remaining dimensionful MSSM parameters are the higgsino mass $\mu$, the
Higgs mass mixing parameter $B_\mu$, and the trilinear $a$-parameters. We assume
that a $M_{\rm P}$-sized $\mu$ at the renormalisable level is absent, and focus
on how a suitable effective $\mu$ term is induced after supersymmetry breaking.
It is well known that no $\mu$ and $B_\mu$ terms are generated in minimal gauge
mediation. Gravity mediation, however, does generically generate them: For
instance, terms of the form
\be\label{mubmu}
{\cal L}=\int d^4\theta\,\frac{X^\dag}{M_{\rm P}}\,H_u H_d+\int
d^4\theta\,\frac{X^\dag X}{M_{\rm P}^2}\,H_u H_d\hc
\ee
induce $\mu= \ol{F}/M_{\rm P}$ and $B_\mu= \left|F/M_{\rm P}\right|^2$
\cite{Giudice:1988yz}.

Trilinear $a$-terms in gauge mediation arise only at higher loop order, and are
thus suppressed with respect to the gaugino masses. Provided that the number of
messengers is not large enough to compensate for this loop suppression (which it
never is in the models we are concerned with), the trilinear terms are
predominantly generated by gravity mediation,  $a\sim F/M_{\rm P}$.

The gravitino mass $m_{3/2}$, finally, is given by
\be
m_{3/2}=\frac{F}{\sqrt{3}M_{\rm P}}\,
\ee
in Minkowski vacua. It is thus of the order of the gravity-mediated soft masses
as usual. 

If gauge-mediation is to be the dominant source for gaugino and scalar soft
masses, there should be a sizeable number of messenger fields. Messenger
multiplicities are bounded from above in conventional low-scale gauge mediation,
since too many messengers would cause the gauge couplings to run into a Landau
pole below the GUT scale. But in high-scale models like ours this is clearly not
a concern, as the messengers decouple already at the GUT-scale.

What can instead be a problem is the different scaling behaviour of scalar and
gaugino masses with the messenger number. For illustration, consider a unified
model containing $N_5$ pairs of ${\bf 5}\oplus\ol{\bf 5}$ messengers with
GUT-scale masses. The gaugino masses grow as $N_5$, while the scalar masses grow
as $\sqrt{N_5}$. For $N_5$ large enough that gauge mediation dominates,
$N_5\gtrsim 5$ say, it becomes difficult to achieve realistic electroweak
symmetry breaking. More precisely, if $M_{3}\gg m_{H_u}$ at the GUT scale, then
$m_{H_u}^2$ will run down very quickly towards the low scale. This effect is
caused mainly by loops of third generation squarks, whose masses in turn are
driven large by gluino loop corrections. Thus $m_{H_u}^2$ at the electroweak
scale will be large and negative, and should be compensated for by a similarly
large $|\mu|^2$, as can best be seen from the relation
\be\label{zmass}
-\frac{M_Z^2}{2}\simeq|\mu|^2+m_{H_u}^2\,,
\ee
valid for $\tan\beta\gtrsim 5$. Eq.~\eqref{zmass} clearly requires a large
cancellation on the RHS in order to reproduce the measured value of $M_Z\approx
91$ GeV. In our scenario $\mu$ is suppressed at the GUT scale, and will not
change significantly through renormalisation group running. Therefore a large
negative $m_{H_u}^2$ cannot be cancelled in Eq.~\eqref{zmass}, so the gluino
mass should not be large to begin with.

This problem can be solved in models where the messengers do not form complete
GUT multiplets, and whose gaugino masses are therefore non-universal.
Specifically, the messenger sector should comprise a large number of weak
doublet messengers and comparatively few colour triplet pairs; this will lead to
larger (positive) $m_{H_u}^2$ and smaller $M_3$ at the high scale. Again, in
usual low-scale gauge mediation such a model would be incompatible with gauge
coupling unification, but if the messengers decouple around the GUT scale, they
will barely influence the evolution of the gauge couplings. One might object
that, in the framework of conventional 4D GUTs, the essential feature of grand
unification is precisely to have all states in complete GUT multiplets at and
above the GUT scale. However, 
as pointed out in the Introduction, this is not the case in the framework of
orbifold GUTs
 \cite{Kawamura:1999nj,Hall:2001pg,Hebecker:2001wq,Asaka:2001eh}, or related
heterotic orbifold models
\cite{Kobayashi:2004ud,Forste:2004ie,Buchmuller:2005jr,Buchmuller:2006ik}. They,
on the contrary, generically predict various states in split representations.
Split multiplets are also common within a different class of string
constructions, namely, F-theory GUTs \cite{Donagi:2008ca,Beasley:2008dc}. In
F-theory the GUT group group is broken via hypercharge flux, which leads to
exotic states in incomplete GUT multiplets \cite{Marsano:2009gv}. In fact it has
been argued that such states might even be required in order to be consistent
with the observed gauge coupling
unification \cite{Blumenhagen:2008aw,Marsano:2009gv}.

Our main example is motivated by the heterotic string construction of
\cite{Buchmuller:2005jr, Buchmuller:2006ik}. It exhibits not only messengers in
split representations, but even a pattern of precisely the sort that is
favourable for electroweak symmetry breaking according to the above discussion:
There are more weak doublets than colour triplets, hence the gluino mass is
suppressed and the soft Higgs masses are enhanced.

One of the main reasons why low-scale gauge mediation is considered to be
attractive is that it generates flavour-blind soft terms. This is still the case
in our scenario, but as opposed to the low-scale case, here the gravity-mediated
contributions to soft terms are sizeable. They will generically induce
unacceptably large flavour-changing neutral currents.\footnote{See
e.g.~\cite{Feng:2007ke,Nomura:2007ap,Hiller:2008sv} for recent studies of the
flavour problem in hybrid gauge-gravity mediation.} To suppress flavour
violation in the gravity-mediated soft terms, an additional mechanism is
necessary -- for instance, wave-function localisation or horizontal symmetries.

A fairly robust prediction of our scenario is large $\tan\beta$: The $B_\mu$
parameter is induced by gravity mediation and therefore suppressed with respect
to $m_{H_d}^2$. Thus 
\be
\tan\beta\simeq\frac{m_{H_u}^2+m_{H_d}^2+2|\mu|^2}{2\,B_\mu}
\ee
is large at the high scale. This remains true after running to the electroweak
scale, since $B_\mu$ cannot grow large if $\mu$ is small, and since $m_{H_d}^2$
cannot shrink significantly unless $\tan\beta$ is large.

A related feature is a scalar $\tau$ lepton which is relatively light compared
to the remaining squarks and sleptons. In fact we will see that, when we discuss
concrete models, squark and slepton masses will be around at least half a TeV,
while the higgsino-like neutralinos and chargino will have masses around
$100-200$ GeV. The lighter $\tilde\tau$ can have a mass somewhere in between:
Since $\tilde\tau_R$ only has hypercharge gauge interactions, the RG evolution
of $m_{\tilde\tau_R}$ will be mainly driven by the $\tau$ Yukawa coupling to the
Higgs sector, which is large at large $\tan\beta$ and drives  $m_{\tilde\tau_R}$
down.

Other characteristics of the low-energy spectrum include near-degenerate masses
for the light neutralinos and chargino (with mass differences of the order of a
few GeV), and a similar mass degeneracy for the heaviest neutralino and the
heavier charginos. This is because the former are almost completely higgsinos,
while the latter are almost completely Winos. Furthermore, the heavier neutral
Higgs scalar is almost degenerate with the pseudoscalar Higgs and close in mass
to the charged Higgs, since we are in an extreme decoupling limit.

It is worthwile to briefly compare our approach with the scenario dubbed ``sweet
spot supersymmetry'' in \cite{Ibe:2007km}, which has also been argued to be
realised in F-theory \cite{Marsano:2008jq}, and where the MSSM soft parameters and
$\mu$ are also generated from a combination of high-scale gauge mediation and
higher-dimensional operators. The authors of \cite{Ibe:2007km} propose
gauge-mediated SUSY breaking with $m_{3/2}\approx 1$ GeV, roughly corresponding
to $F\approx(10^{9}\,\text{GeV})^2$ and $M_{\rm m}\approx 10^{13}$ GeV. Furthermore
they assume that $\mu$ and $B_\mu$ are generated by  higher-dimensional
operators whose suppression scale $\Lambda$ is below the Planck scale. For
instance, the Higgs fields could couple directly to the hidden sector, which
would induce the operators
\be\label{sweetspot}
{\cal L}=\int d^4\theta\,\frac{X^\dag}{\Lambda}\,H_u H_d\hc+\int
d^4\theta\,\frac{X^\dag X}{\Lambda^2}\left(H_u^\dag H_u+H_d^\dag H_d\right)\,
\ee
after integrating out certain hidden sector states of mass $\Lambda$. For
$\Lambda\approx 10^{16}$ GeV Eq.~\eqref{sweetspot} gives an electroweak-scale
$\mu$ term, as well as sizeable contributions to the Higgs soft masses. Such
models may thus be viewed as standard gauge mediation models with two unusual
features: The $\mu$ problem is solved by directly coupling the Higgs fields to
the hidden sector, and the gravitino is a viable dark matter candidate. By
construction, gravity mediation proper (i.e.~visible sector soft terms induced
by $M_{\rm P}$-suppressed operators) never plays a role. Indeed similar models
could be constructed with lower $m_{3/2}$ and correspondingly lower $\Lambda$,
$F$, and $M_{\rm m}$, if one gives up the requirement that the gravitino should
be dark matter. 

Our scenario, by contrast, requires the introduction of just a single new scale:
the scale of supersymmetry breaking $\sqrt{F}\approx 10^{10}$ GeV. The messenger
mass is given by the GUT scale or compactification scale, and the suppression
scale for the $\mu$ term is fixed to be $M_{\rm P}$. We consider this quite
appealing from the model-building point of view. However, in our restricted
framework there is naturally less freedom to evade experimental constraints by
simply adjusting the scales. For instance, as previously stated, in our scenario
the flavour problem is not automatically solved, but requires some extra
mechanism.  

Note also that our models are very different from hybrid gauge-gravity mediation
with anomalous $\U 1$s where all messenger fields are in complete GUT multiplets 
\cite{Dudas:2007nz}. While such models also allow for gravitino dark matter, this 
requires all other superparticle masses to be at least around a TeV 
\cite{Cerdeno:2009ns}.

Concerning the superparticle spectrum, our models share some characteristic 
features with ``lopsided gauge mediation'' \cite{DeSimone:2011va}, which also 
predicts light higgsinos along with coloured states heavier than a TeV. However, 
in contrast to our models, the sleptons and the bino still tend to be relatively 
light in the models of \cite{DeSimone:2011va}, while the pseudoscalar Higgs is 
extremely heavy. The theoretical framework is, of course, also very different,
lopsided gauge mediation essentially being pure gauge mediation with certain 
direct Higgs-messenger couplings.

\section{Models}\label{models}

\subsection{A toy model}\label{toymodel}

Let us first present a model whose purpose it is to illustrate our scenario in
the simplest possible way, without being motivated by any specific UV
completion. Consider $N_5$ pairs of messengers in the ${\bf 5}\oplus\ol{\bf 5}$
of $\SU 5$, and $N_2$ pairs of messengers in the ${\bf 2}_{1/2}\oplus{\bf
2}_{-1/2}$ of $\SU{2}_L\times\U{1}_Y$. We take all messenger masses to be equal,
$M_{\rm m}= 10^{16}$ GeV, and $F=(2\cdot 10^{10}\text{ GeV})^2$, corresponding
to $m_{3/2}= 100$ GeV. Furthermore we set
\be\begin{split}
\mu&\sim\frac{F}{M_{\rm P}},\\
A_0&\sim\frac{F}{M_{\rm P}},\\
B_\mu&\sim\frac{F^2}{M_{\rm P}^2}
\end{split}
\ee
at $M_{\rm GUT}$, to account for these terms being generated by gravity
mediation. Here $A_0$ is, as usual, related to the trilinear $a$-parameters by
\be
a_{u,d,l}=A_0\,y_{u,d,l}
\ee
with $y_{u,d,l}$ the Yukawa matrices.

The gauge-mediated contributions to the gaugino masses are
\be\begin{split}
M_3&=\frac{g^2}{16\pi^2}N_5\,\frac{F}{M_{\rm m}}\,, \\
M_2&=\frac{g^2}{16\pi^2}\left(N_5+N_2\right)\,\frac{F}{M_{\rm m}}\,,\\
M_1&=\frac{g^2}{16\pi^2}\left(N_5+\frac{3}{5} N_2\right)\,\frac{F}{M_{\rm m}}\,.
\end{split}
\ee
For the scalar masses, we obtain
\be\begin{split}
m_Q^2&=\left(\frac{g^2}{16\pi^2}\right)^2\left(\frac{F}{M_{\rm
m}}\right)^2\,\left(\frac{21}{5}N_5+\frac{38}{25}N_2\right)\,,\\
m_U^2&=\left(\frac{g^2}{16\pi^2}\right)^2\left(\frac{F}{M_{\rm
m}}\right)^2\,\left(\frac{16}{5}N_5+\frac{8}{25}N_2\right)\,,\\
m_D^2&=\left(\frac{g^2}{16\pi^2}\right)^2\left(\frac{F}{M_{\rm
m}}\right)^2\,\left(\frac{14}{5}N_5+\frac{2}{25}N_2\right)\,,\\
m_L^2=m_{H_u}^2=m_{H_d}^2&=\left(\frac{g^2}{16\pi^2}\right)^2\left(\frac{F}{M_{
\rm m}}\right)^2\,\left(\frac{9}{5}N_5+\frac{42}{25}N_2\right)\,,\\
m_E^2&=\left(\frac{g^2}{16\pi^2}\right)^2\left(\frac{F}{M_{\rm
m}}\right)^2\,\left(\frac{6}{5}N_5+\frac{18}{25} N_2\right)\,.
   \end{split}
\ee

Neglecting gravity-mediated contributions to scalar soft masses and gaugino
masses altogether, it is possible to obtain realistic electroweak symmetry
breaking e.g.~with the following parameters:
\be\begin{split}\label{toyparams}
N_5&=6\\
N_2&=10\\
\mu&=145\,{\rm GeV},\,\\
A_0&=145\,{\rm GeV},\,\\
B_\mu&=(223\,{\rm GeV})^2\,.
\end{split}
\ee
The gauge-mediated soft masses at $M_{\rm GUT}=1.2\,\cdot\,10^{16}$ GeV are then
as in Table \ref{toytable}. The low-energy particle spectrum as computed with
\verb!SOFTSUSY! \cite{Allanach:2001kg} is shown in Table \ref{toytable2}. As
expected $\tan\beta$ turns out to be rather large, $\tan\beta=48$. Generally the
spectrum is much as anticipated in Section \ref{softterms}. There is a Standard
Model-like Higgs with its mass just above the LEP bound. The higgsino-like
neutralinos and chargino are light and almost degenerate in mass. Except for a
relatively light $\tilde\tau$, all other states have masses above 600 GeV and up
to more than 2 TeV. The heavier Higgs (pseudo)scalars have very similar masses,
and also the heavier chargino is nearly degenerate with the heaviest
neutralino.

\begin{table}
\begin{center}
\begin{tabular}{|c|c|c|}\hline mass parameter & value [GeV]\\ \hline\hline
$M_1$ & $1592$\\ 
$M_2$ & $2122$ \\ 
$M_3$ & $796$  \\
\hline 
$m_Q$ & $843$ \\ 
$m_U$ &  $627$ \\ 
$m_D$ & $556$ \\
\hline 
$m_L=m_{H_u}=m_{H_d}$ & $697$  \\ 
$m_E$ & $503$ \\ 
 \hline 
\end{tabular}
\caption{GUT-scale parameters for a toy model with $N_5=6$ pairs of $5$-plet
messengers and $N_2=10$ pairs of doublet messengers.}\label{toytable}
\end{center}
\end{table}

\begin{table}
\begin{center}
\begin{tabular}{|c|c|c|}\hline particle &  mass [GeV]\\ \hline\hline
$h_0$ & $118$\\ \hline
$\chi^0_1$ &  $124$ \\ 
$\chi^\pm_1$ & $127$  \\ 
$\chi^0_2$ & $130$ \\ 
\hline
$\chi^0_3$ &  $691$ \\ 
$\chi^0_4$ & $1724$ \\ 
$\chi^\pm_2$ & $1724$  \\ \hline
$H_0$ & $764$ \\ 
$A_0$ & $765$ \\ 
$H^\pm$ & $770$ \\ \hline
$\tilde g$ & $1792$ \\ \hline
$\tilde\tau_1$ & $211$ \\
other sleptons & $780 - 1550$\\ \hline
squarks & $1090 - 2180$ \\\hline
\end{tabular}
\caption{Low-energy spectrum for the mass parameters of Table \ref{toytable}, 
with $\mu=A_0=145$ GeV and $B_\mu=(223$ GeV$)^2$\,.}\label{toytable2}
\end{center}
\end{table}

\subsection{A model from a heterotic orbifold}\label{hetmodel}

Higher-dimensional orbifold GUTs, or heterotic string compactifications,
naturally contain incomplete GUT multiplets. These can arise from twisted states,
localised at orbifold singularities where only a subgroup of the GUT group
is realised; or from untwisted states whose zero modes are partly projected
out by the orbifold. In addition, there are generally also massless pairs of 
vector-like exotics in complete GUT representations. All of these exotics should 
eventually become massive, leaving only a pair of light Higgs doublets
and the three generations of chiral MSSM matter at low energies. In principle a
limited number of complete GUT multiplets could survive down to energies far
below $M_{\rm GUT}$, without affecting perturbative gauge coupling unification.
Here, however, we assume for simplicity a common mass for all messengers, which 
should then be close to $M_{\rm GUT}$ (see also Section \ref{multi}).

 There are two ways in which the GUT scale can enter in heterotic orbifolds.
First, usually some Standard Model singlets need to take vacuum expectation
values in order to cancel the FI term of an anomalous $\U 1$, whose magnitude is
about a loop factor below $M_{\rm P}$. This introduces the scale $M_{\rm
P}/(16\pi^2)\approx M_{\rm GUT}$ into the scalar potential (similar as in the
related string-inspired models of \cite{Dudas:2007nz}). Second, MSSM gauge
coupling unification motivates considering anisotropic compactifications
\cite{Witten:1996mz, Hebecker:2004ce}, where one or two of the radii of the
internal manifold are large in string units (around $1/M_{\rm GUT}$) while the
others are small (around $1/M_{\rm P}$). One obtains an intermediate effective
description between $M_{\rm GUT}$ and $M_{\rm P}$ in terms of a 5D or 6D
orbifold GUT. In this picture, the compactification of the larger two extra
dimensions breaks the 5D or 6D bulk gauge symmetry to the Standard Model.
Such a large compactification radius may again have its dynamical origin in
the scale of the FI term, depending on the moduli stabilisation mechanism
\cite{Buchmuller:2008cf}.

Our prime example for this Section will be the model of
\cite{Buchmuller:2006ik}.  The massless spectrum contains various chiral
supermultiplets, both Standard Model singlets and fields with Standard Model
charges. Besides the three generations of quarks and leptons, and a massless
pair of Higgs doublets, there are several vector-like exotics which become
massive when some of the singlet fields acquire vevs. The massless vector-like
exotics are listed in Table \ref{messengers}, where we have also indicated the
geometric origin of the zero modes in a 6D orbifold GUT limit.\footnote{In
principle, higher KK modes coming from bulk states will also act as messengers;
a more detailed study should take their contributions into account.} 
\begin{table}
\begin{center}
\begin{tabular}{|c|c|c|c|}\hline field & representation & multiplicity & 6D
origin \\ \hline
$d$ & $({\bf 3},{\bf 1})_{-1/3}$ & $4$ & bulk\\ 
$\tilde d$ & $(\ol {\bf 3},{\bf 1})_{1/3}$ & $4$ & bulk\\ 
$\ell$ & $({\bf 1},{\bf 2})_{1/2}$ & $4$ & bulk \\ 
$\tilde\ell$ & $({\bf 1},{\bf 2})_{-1/2}$ & $4$ & bulk\\ 
$m$ & $({\bf 1},{\bf 2})_{\,0}$ & $8$ & brane\\ 
$s^+$ & $({\bf 1},{\bf 1})_{1/2}$ & $16$ & brane\\ 
$s^-$ & $({\bf 1},{\bf 1})_{-1/2}$ & $16$ & brane \\ \hline
\end{tabular}
\end{center}
\caption{The messenger content of a heterotic orbifold
model.}\label{messengers} 
\end{table}

This messenger content does not lead to realistic electroweak symmetry breaking
when coupled to a hidden sector in the minimal manner as in the model of Section
\ref{toymodel}, with a single goldstino background field and a common mass
scale for all messengers. However, such a minimal setup is inappropriate for the
present model for two reasons. First, the effective messenger-goldstino
couplings (the $\lambda_i$ in Eq.~\eqref{mingmW}) are expected to be different
for different types of messengers. We will comment on possible consequences of
this in Section \ref{multi}. Second, different messengers with their different
geometric origins have distinct transformation properties under discrete
symmetries. In our case the selection rules turn out to be such that there
cannot be a single goldstino multiplet coupling to both bulk and brane fields.
The next simplest option, therefore, is to introduce two SUSY-breaking
background superfields $X_1$ and $X_2$ with couplings
\be
W=X_1\,d\tilde d+X_1\ell\tilde\ell+X_2 mm+X_2 s^+ s^-\,.
\ee
For simplicity we take their expectation values to be equal in the lowest
component, but we do allow their $F$-terms to be distinct:
\be
\vev{X_1}=M_{\rm m}+F_1\,\theta^2,\qquad\vev{X_2}=M_{\rm m}+F_2\,\theta^2\,.
\ee
We can define a goldstino mixing angle $\phi$ by
\be\begin{split}
F_1=F\cos\phi\,,&\quad F_2=F\sin\phi\,,\\
\frac{F}{\sqrt 3 M_{\rm P}}&=m_{3/2}\,.\end{split}
\ee
This gives for the gauge-mediated gaugino masses at the scale $M_{\rm m}$
\be\begin{split}\label{hetgauginos}
M_1&=\frac{g^2}{16\pi^2}\frac{F}{M_{\rm
m}}\left(4\cos\phi+\frac{24}{5}\sin\phi\right),\\  
M_2&=\frac{g^2}{16\pi^2}\frac{F}{M_{\rm m}}\left(4\cos\phi+4\sin\phi\right),\\ 
M_3&=\frac{g^2}{16\pi^2}\frac{F}{M_{\rm m}}\;4\cos\phi\,,
   \end{split}
 \ee
and for the scalar soft masses-squared
\be\begin{split}
m_Q^2&=2\left(\frac{g^2}{16\pi^2}\right)^2\left(\frac{F}{M_{\rm
m}}\right)^2\,\left(\frac{287}{50}+\frac{133}{50}\cos 2\phi\right)\,,\\
m_U^2&=2\left(\frac{g^2}{16\pi^2}\right)^2\left(\frac{F}{M_{\rm
m}}\right)^2\,\left(\frac{96}{25}+\frac{64}{25}\cos 2\phi\right)\,,\\
m_D^2&=2\left(\frac{g^2}{16\pi^2}\right)^2\left(\frac{F}{M_{\rm
m}}\right)^2\,\left(\frac{74}{25}+\frac{66}{25}\cos 2\phi\right)\,,\\
m_L^2&=m_{H_u}^2=m_{H_d}^2=2\left(\frac{g^2}{16\pi^2}\right)^2\left(\frac{F}{M_{
\rm m}}\right)^2\,\left(\frac{183}{50}-\frac{3}{50}\cos 2\phi\right)\,,\\
m_E^2&=2\left(\frac{g^2}{16\pi^2}\right)^2\left(\frac{F}{M_{\rm
m}}\right)^2\,\left(\frac{66}{25}-\frac{6}{25}\cos 2\phi\right)\,.
   \end{split}
 \ee

In a 6D orbifold GUT picture, the messenger mass $M_{\rm m}$ should take a value
around the scale of the two larger radii, which is slightly below the 4D
unification scale. For concreteness we set $M_{\rm m}=5\cdot 10^{15}$ GeV. Then
$g$ is still approximately given by the unified gauge coupling at the GUT scale,
$g\approx 0.7$. We again choose $F=(2\cdot 10^{10}\text{ GeV})^2$, so the
gravitino mass is $m_{3/2}=100$ GeV.

\begin{figure}
\begin{center}
\includegraphics[width=.7\textwidth]{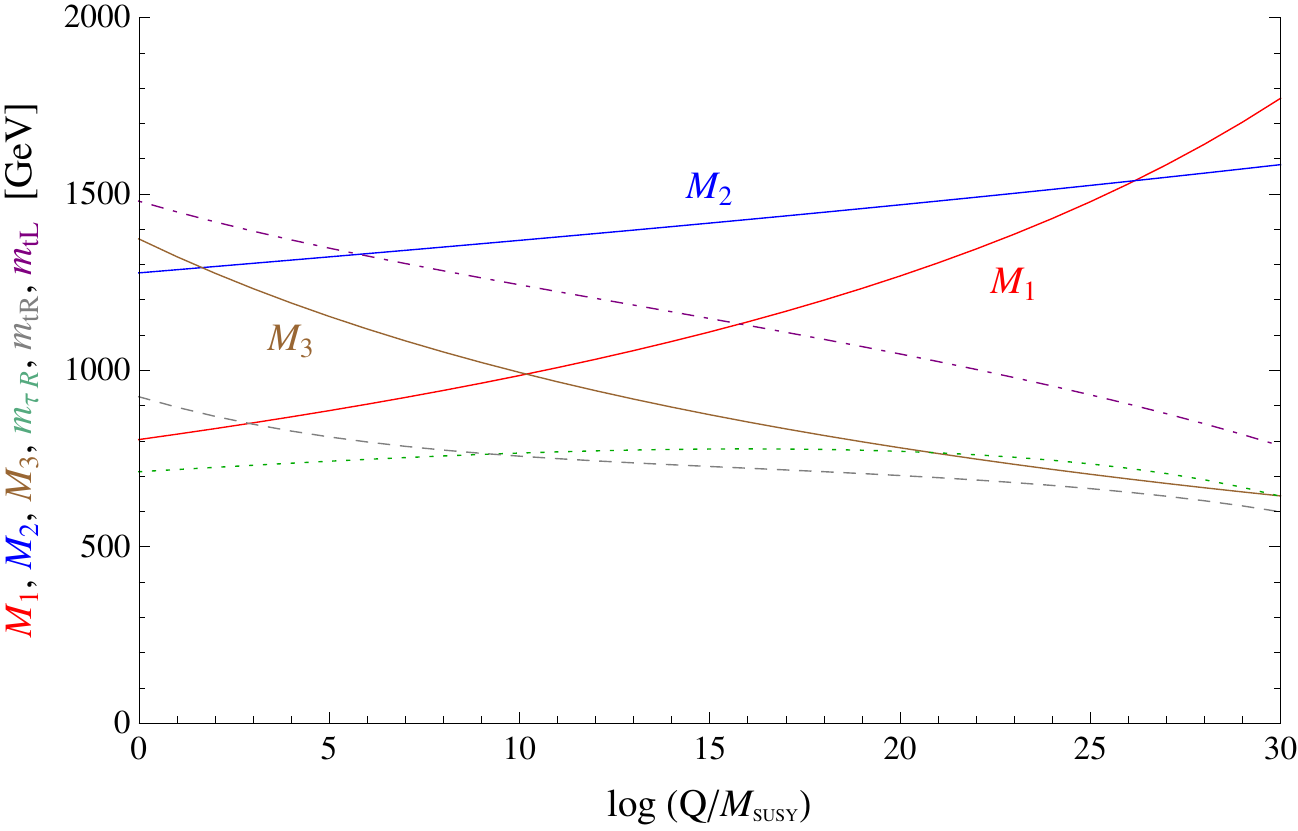}
\end{center} 
\caption{Two-loop renormalisation group evolution of gaugino masses and some
selected sfermion masses for the model of Section \ref{hetmodel}. The horizontal
axis ranges between $M_{\rm SUSY}=1.17$ TeV on the left and the messenger scale
$M_{\rm m}=5\cdot 10^{15}$ GeV on the right. Bino, wino and gluino masses are
shown in red, blue, and brown. The green dotted curve shows the right-handed
stau soft mass; the grey dashed and purple dot-dashed curves are the
right-handed and left-handed stop soft masses.}\label{gaugscplots}
\end{figure}

\begin{figure}
\begin{center}
\includegraphics[width=.7\textwidth]{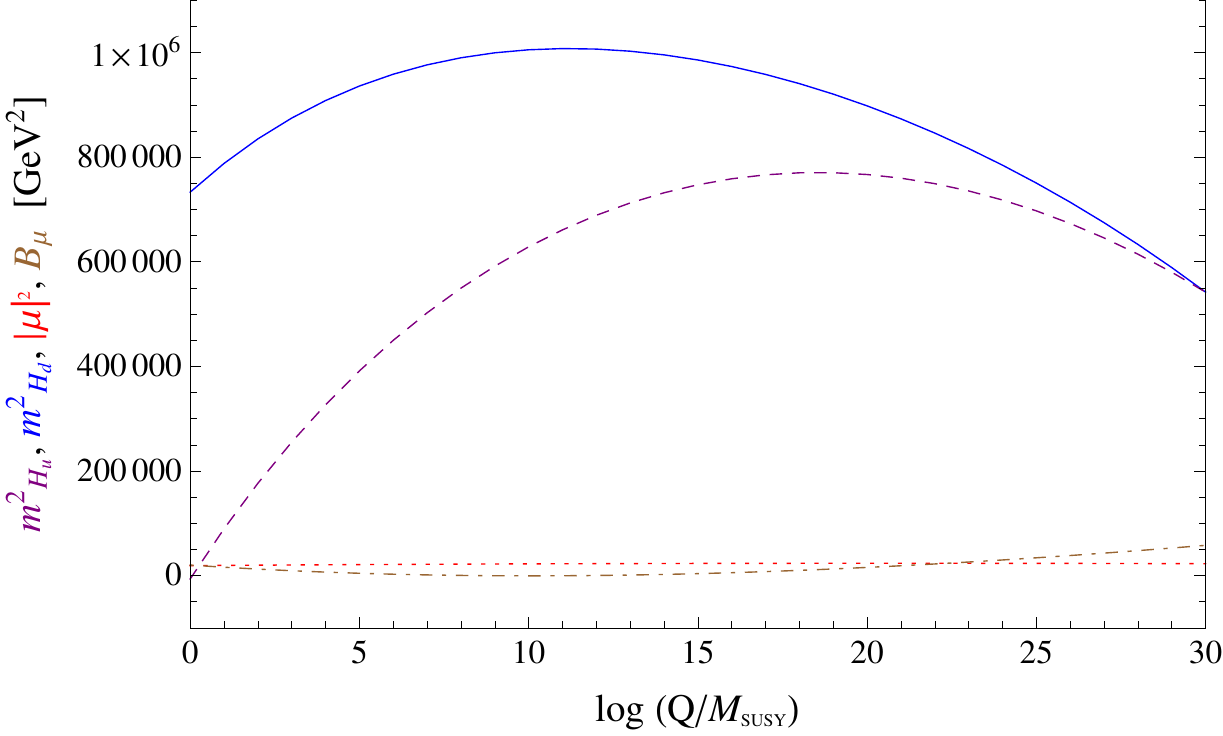}
\end{center} 
\caption{Renormalisation group evolution of Higgs mass parameters for the model
of Section \ref{hetmodel}. The horizontal axis is as in Fig.~\ref{gaugscplots}.
The blue solid curve and the purple dashed curve show $m_{H_d}^2$ and
$m_{H_u}^2$. The red dotted curve shows $|\mu|^2$,  and the brown dot-dashed
curve shows $B_\mu$. }\label{higgsplots}
\end{figure}

To investigate this model, we now take into account both gauge-mediated and
gravity-mediated contributions to soft terms. To account for gravity mediation,
we add a universal piece $m_0$ to the gaugino masses of
Eqns.~\eqref{hetgauginos}, and to scalar masses in quadrature. Choosing
\be
m_0=150\text{ GeV}= 0.9\,\frac{F}{M_{\rm P}}
\ee
and setting
\be\begin{split}\label{hetparams}
\tan\phi&=1.9\,,\\
\mu&=m_0\,,\\
A_0&=m_0\,,\\
B_\mu&=\left(1.6\,m_0\right)^2=(240\,{\rm GeV})^2\,,
\end{split}
\ee
the high-scale mass parameters are as listed in Table \ref{hettable}. These
parameter values lead to realistic electroweak symmetry breaking at
$\tan\beta=41$ and a low-energy spectrum as in Table \ref{hettable2}. The
renormalisation group evolution of some selected parameters between the
messenger scale and the TeV scale is depicted in Figs.~\ref{gaugscplots} and
\ref{higgsplots}. In these plots, $M_{\rm SUSY}=1.17$ TeV is the scale where the
Higgs potential is minimised, chosen to be the geometric mean of the stop
masses. Fig.~\ref{higgsplots} clearly shows the discrepancy between the
gravity-mediated $\mu$ and $B_\mu$ on the one hand, and the gauge-mediated
$m_{H_d}^2$ and $m_{H_u}^2$ on the other hand. Through its renormalisation group
evolution, $|m_{H_u}^2|$ eventually becomes comparable with $|\mu|^2$ at
$M_{\rm SUSY}$, to produce an electroweak symmetry breaking scale which is of
the order of the gravity-mediated terms.

\begin{table}
\begin{center}
\begin{tabular}{|c|c|c|}\hline mass parameter & value [GeV]\\ \hline\hline
$M_1$ & $1771$\\ 
$M_2$ & $1583$ \\ 
$M_3$ & $644$  \\
\hline 
$m_Q$ & $786$ \\ 
$m_U$ &  $599$ \\ 
$m_D$ & $478$ \\
\hline 
$m_L=m_{H_u}=m_{H_d}$ & $736$  \\ 
$m_E$ & $643$ \\ 
 \hline 

\end{tabular}
\caption{GUT-scale parameters for the heterotic model of Section \ref{hetmodel},
with $\cos\phi=0.466$ and a universal gravity-mediated piece $m_0=150$
GeV.}\label{hettable}
\end{center}
\end{table}

\begin{table}
\begin{center}
\begin{tabular}{|c|c|c|}\hline particle &  mass [GeV]\\ \hline\hline
$h_0$ & $117$\\ \hline
$\chi^0_1$ &  $137$ \\ 
$\chi^\pm_1$ & $140$  \\ 
$\chi^0_2$ & $144$ \\ 
\hline
$\chi^0_3$ &  $799$ \\ 
$\chi^0_4$ & $1296$ \\ 
$\chi^\pm_2$ & $1296$  \\ \hline
$H_0$ & $856$ \\ 
$A_0$ & $857$ \\ 
$H^\pm$ & $861$ \\ \hline
$\tilde g$ & $1453$ \\ \hline
$\tilde\tau_1$ & $713$ \\
other sleptons & $910 - 1290$\\ \hline
squarks & $950 - 1750$ \\\hline
\end{tabular}
\caption{Low-energy spectrum for the mass parameters of Table \ref{hettable}, 
with $\mu=A_0=150$ GeV and $B_\mu=(240$ GeV$)^2$\,.}\label{hettable2}
\end{center}
\end{table}

It is interesting that in this model the lighter $\tilde\tau$ is still
relatively heavy, especially when compared to the model of Section
\ref{toymodel}. This is due to the smaller $\tan\beta$ in the current setup, as
well as to the presence of a large number of hypercharged messengers which
increase the GUT-scale value of $m_E$ relative to the other masses. Otherwise
the spectrum is qualitatively very similar to that in Section \ref{toymodel}.

Regarding the fundamental input parameters, note that the choice
$\phi>\frac{\pi}{4}$ (i.e.~$\tan\phi>1$) serves to further suppress $M_3$ with
respect to $M_2$ and $M_1$. As discussed in Section \ref{softterms}, a too large
gluino mass would be in conflict with electroweak symmetry breaking. This
problem is now solved in our model in a twofold way: First, there are more
messengers charged under $\SU{2}_L\times\U{1}_Y$ than under $\SU{3}_C$; and
second, the former couple somewhat more strongly to the hidden sector than the
latter.

\subsection{Comments on multiple messenger scales}\label{multi}

In the previous Section we have assumed a common messenger mass scale $M_{\rm
m}$ for all messenger fields. It would be more realistic to take into account
that different messenger fields will, in fact, decouple at different scales.
More precisely, in the heterotic orbifold model the messenger superpotential is,
to quadratic order in the messengers $\Sigma_i,\,\widetilde\Sigma_i$,
\be\label{messW}
W=\sum_i M_{\rm P}\,{\cal P}_i\left(\frac{S_I}{M_{\rm
P}}\right)\Sigma_i\widetilde\Sigma_i\,.
\ee
Here the ${\cal P}_i$ are certain polynomials of degree $>1$, and the $S_I$ are
Standard Model singlet fields. The dimensionless coefficients entering the
${\cal P}_i$ are unknown (but could in principle be computed from the worldsheet
CFT amplitudes of the underlying string model). The expectation values of the
$S_I$ are also unknown, but a naive estimate gives $\vev{S_I}\approx M_{\rm
GUT}$, which would be the correct order of magnitude to cancel a $M_{\rm
GUT}$-sized FI term. Some of the $S_I$ should also acquire $F$-term vevs from
coupling to the hidden sector. Denoting by $n_i$ the degree of the lowest
non-vanishing monomial in ${\cal P}_i$, and assuming that all unknown
coefficients are generic and of order unity, one obtains an effective
superpotential resembling Eq.~\eqref{mingmW}:
\be
W=\sum_i\lambda_i\,X_i\Sigma_i\widetilde\Sigma_i\,,
\ee
where $\lambda_i\sim\left(\frac{M_{\rm GUT}}{M_{\rm P}}\right)^{n_i-1}$, and
$\vev{X_i}\sim M_{\rm GUT}+F\theta^2$.

It is evident that, when taking these naive arguments seriously, messengers
which couple to higher powers of the $S_I$ decouple at lower scales, while their
gauge-mediated contribution to soft masses remains similar. Strictly speaking,
however, our reasoning can at most serve as a rough indication that the
messenger scales in this model can be hierarchically different. To treat this
matter quantitatively, one would have to undertake the formidable task of
computing the unknown couplings between the light modes, and subsequently
minimising the full scalar potential.

Let us therefore merely point out how multiple messenger scales could help to
render the model viable phenomenologically. Messengers in incomplete GUT
multiplets, such as the $m$ and $s^{\pm}$ fields, should still decouple around
the GUT scale in order not to interfere with gauge coupling unification. On the
other hand, the $d$, $\tilde d$, $\ell$ and $\tilde\ell$ fields happen to fall
into complete ${\bf 5}\oplus\ol{\bf 5}$ pairs (even though they do not originate 
from the same GUT multiplets in the string model), so their 
decoupling scale can be much lower.\footnote{Indeed the $n_i$ are found to be 
large precisely for these fields in  \cite{Buchmuller:2006ik}, but as already 
stated it would be premature to infer that they are correspondingly lighter.} 
This can alleviate the problem which we pointed out in Section \ref{softterms}, 
namely, that large gluino masses tend to spoil electroweak symmetry breaking via 
their effect on renormalisation group running. In our model gauge-mediated gluino 
masses are induced only by $d$ and $\tilde d$ messengers. Therefore, they will be 
generated at low scales (if the $d$ and $\tilde d$ decouple at low energies), and 
hence will have less influence on the running of the Higgs mass parameters.

\subsection{Naturalness}

From Fig.~\ref{higgsplots} it is evident that some amount of fine-tuning is
required in our model in order to reproduce realistic electroweak symmetry
breaking. Namely, the input parameters have to be chosen such that $|m_{H_u}^2|$
is very small at $M_{\rm SUSY}$ for Eq.~\eqref{zmass} to be satisfied.

A quantitative measure of fine-tuning \cite{Barbieri:1987fn} is the maximum
sensitivity of $M_Z$ with respect to variations of any of the continuous
parameters,
\be\label{naturalness}
c=\underset{a\in\,\{{\rm Parameters}\}}{\rm max}\;\left|\frac{\partial\log
M_Z}{\partial\log a}\right|\,.
\ee 
Taking the set of parameters to be the MSSM parameters of Table \ref{hettable},
along with $\mu$ and $B_\mu$, we find for the models of Section \ref{toymodel}
and Section \ref{hetmodel} $c$-values of around $c\sim 200$, so the models could
be considered to be fine-tuned at a level comparable with the CMSSM in its
remaining viable parameter regions.

However, one should bear in mind that for our class of models, the dimensionful 
MSSM parameters are derived quantities and cannot be varied independently. A 
more appropriate set of fundamental parameters for the model of Section
\ref{toymodel} would thus be $\{F/M_{\rm m},\,A_0,\,\mu,\,\sqrt{B_\mu}\}$, in
addition to the discrete messenger numbers which do not enter in the definition
of $c$, Eq.~\eqref{naturalness}. Computing the $c$-value with respect to these
parameters, one obtains $c\sim 20$, which is a significant improvement (albeit
still not quite ${\cal O}(1)$). This can be understood as follows:
Gauge-mediated contributions to soft terms are large, and are governed by
discrete parameters, while gravity-mediated contributions are small, and are
determined by continuous parameters. Furthermore, if $\mu$ is small as in our
case, then $M_2$ and $M_3$ are the MSSM soft terms to which the electroweak
scale is by far the most sensitive. The fine-tuning is considerably reduced once
we fix the ratio of the gauge-mediated contributions to $M_2$ and $M_3$ by some
suitable choice of messenger numbers. Variations of the gravity-mediated terms
do not have a large impact as long as gravity mediation is subdominant, and
varying the overall scale of gauge mediation leads to a simultaneous change in
both $M_2$ and $M_3$, the effects of which can (at least partially) cancel. The
overall picture is similar to the one advocated in \cite{Horton:2009ed}, where
it was recently found that, quite generally, certain choices of non-universal
gaugino mass ratios can help to reduce fine-tuning. 

For the model of Section \ref{hetmodel} the fundamental continuous parameters
are $\{F_1/M_{\rm m},\,F_2/M_{\rm m},\,m_0,\,\sqrt{B_\mu}\}$. Since here the
gauge-mediated contributions to $M_2$ and $M_3$ still depend on a continuous
parameter (the ratio $F_1/F_2$, or equivalently the goldstino angle $\phi$), the
$c$-value with respect to this parameter set remains high.

Beyond these simple considerations, the naturalness question should probably be
best left to the framework of a concrete UV completion which more rigorously
defines the set of independent parameters. Of course, ultimately all parameters
in the String Landscape will be discrete, so in that context even the entire
notion of fine-tuning and naturalness here appears questionable.

\section{Cosmology and phenomenology}

A gravitino LSP with a mass of ${\cal O}(100)$ GeV is an interesting dark 
matter candidate. At high reheating temperatures, as required by thermal
leptogenesis, thermal production of gravitinos yields the observed dark matter 
abundance for typical gluino masses. A potential problem, however, are the 
severe constraints from BBN. The NLSP is long-lived, and its late decays  
inject highly energetic particles into the plasma after nucleosynthesis. These 
will destroy the newly formed nuclei and thus distort the successfully 
predicted light element abundances. 

In the analysis of \cite{Bolz:1998ek} this problem could be avoided for a
sufficiently short lived higgsino NLSP since at that time the BBN constraints 
only imposed the upper bound on the higgsino abundance 
$\Omega_{\tilde{h}} h^2 \lesssim 8\cdot 10^{-3}$ for lifetimes 
$\tau_{\tilde{h}} \lesssim 2\cdot 10^6~\mathrm{s}$ \cite{Ellis:1990nb}. 
For a higgsino NLSP 
in the mass range $80~\mathrm{GeV} < m_{\tilde{h}} < 300~\mathrm{GeV}$
the above BBN bound on the higgsino abundance is satisfied due to the effect 
of coannihilations \cite{Mizuta:1992qp,Edsjo:1997bg} and the lifetime 
constraint can be satisfied for gravitino masses below $100~\mathrm{GeV}$.
Hence, a consistent picture of leptogenesis, gravitino dark matter and 
nucleosynthesis could be obtained.

The present BBN bounds on NLSP abundances and lifetimes are much more 
stringent. In the case of dominant hadronic NLSP decays and lifetimes
$\tau_{\mathrm{NLSP}} \gtrsim 10^8~\mathrm{s}$ one finds the upper  
bounds \cite{Kawasaki:2004yh,Jedamzik:2006xz}
\begin{align}
\Omega_{\mathrm{NLSP}} h^2 &\lesssim 1\cdot 10^{-4} \quad 
\mathrm{from}\quad ^2\mathrm{H}\ , \label{BBN1}\\
\Omega_{\mathrm{NLSP}} h^2 &\lesssim 3\cdot 10^{-5} \quad 
\mathrm{from}\quad ^3\mathrm{He}\ \label{BBN2}. 
\end{align}
A detailed analysis for a general neutralino NLSP \cite{Covi:2009bk} has shown 
that, except for special points in parameter space, these constraints can only 
be satisfied for rather short NLSP lifetimes, 
$\tau_{\mathrm{NLSP}} \lesssim 10^2 - 10^3~\mathrm{s}$. For a gravitino mass
of $100~\mathrm{GeV}$ this requires NLSP masses in excess of $2~\mathrm{TeV}$.

In our model the dominant NLSP decays are the 2-body decay into gravitino and
photon, and the 3-body decay into gravitino and hadrons via a virtual 
$Z$-boson. For large $\mathrm{tan}\beta$ the corresponding decay widths are 
given by \cite{Covi:2009bk}
\begin{align}
\Gamma(\chi^0_1\into \psi_{3/2}~\gamma) &\simeq
\frac{\epsilon^2_{h\gamma}}{48\pi M_P^2}\frac{(m_{\chi^0_1})^5}{m_{3/2}^2}
\left(1-\left(\frac{m_{\chi^0_1}}{m_{3/2}}\right)^2\right)^3
\left(1+3\left(\frac{m_{\chi^0_1}}{m_{3/2}}\right)^2\right)\ ,\\
\Gamma(\chi^0_1\into\psi_{3/2}~\text{had}) &\simeq
\frac{r_{\text{had},b\bar b}}{96 (4\pi)^3 M_P^2} 
\frac{(m_{\chi^0_1})^7}{m_{3/2}^2}\frac{g_2^2}{M_Z^2\,c^2_w}
\left(1-\frac{4}{3} s_w^2+\frac{8}{9} s_w^4\right)\ ;   
\end{align}
here $c_w = \cos\theta_w$, $s_w = \sin\theta_w$, $\theta_w$ is the weak mixing 
angle, $r_{\text{had},b\bar b} = 
{\rm BR}(Z\into\text{hadrons})/{\rm BR}(Z\into b\bar b)$, and
\begin{align}
\epsilon_{h\gamma} = -
\frac{s_w c_w M_Z (M_2-M_1)}{\sqrt{2}\,M_1 M_2}
\end{align}
is the higgsino-photino mixing angle. In our model 
$\epsilon_{h\gamma} = -0.013$. Because of this small mixing
the hadronic decays dominate. We find
\begin{align}
\Gamma^{-1}(\chi^0_1\into\psi_{3/2}~\text{had})\simeq 2\cdot 10^{11}~{\rm s}\ ,
\quad 
\Gamma^{-1}(\chi^0_1\into \psi_{3/2}~\gamma) \simeq 3\cdot 10^{12}~{\rm s}\ .
\end{align}

The $\chi^0_1$ relic abundance in our model is predicted to be rather small. 
This is because $\chi^0_1$ is nearly degenerate with $\chi^\pm_1$ in mass, so 
the $\chi^0_1$ can efficiently coannihilate with the chargino 
\cite{Mizuta:1992qp}. Using the \verb!micrOMEGAs! code \cite{Belanger:2006is}, 
we find for the spectrum of Section \ref{hetmodel} a neutralino relic 
abundance of
\be\label{reldens}
\Omega_{\chi^0_1} h^2=3.2\cdot 10^{-3}\ .
\ee 
This is four orders of magnitude smaller than a typical bino relic density,
but it still exceeds the BBN bounds (\ref{BBN1}), (\ref{BBN2}) by one to two
orders of magnitude. This may be remedied in several ways, for instance by 
introducing small R-parity violating couplings \cite{Buchmuller:2007ui} or by 
additional entropy production before nucleosynthesis \cite{Hasenkamp:2010if}. 
Note also that, because of the small $\chi^0_1$ relic density of 
Eq.~\eqref{reldens}, the lightest neutralino is not a possible dark matter 
candidate.

Our model predicts intriguing and distinctive signatures for the LHC 
experiments. The low energy particle spectrum contains only three fermions 
in addition to the Standard Model particles, two of them neutral and one 
charged, almost mass degenerate and close to the Higgs boson mass.
The lightest neutralino will decay mostly outside the detector, even if 
R-parity is broken \cite{Bobrovskyi:2010ps}. The heavier fermions decay
into the next lighter one and a few hadrons via a virtual W-boson. The near
mass degeneracy leads to rather long decay lengths of 
$c\tau_\chi={\cal O}( 1\,\mu\rm{m})$. Energetic decaying higgsinos will lead to events
with missing energy accompanied by low-momentum jets and leptons. These 
signatures, the production of charginos 
and neutralinos in cascades and Drell-Yan processes, and the predictions 
for rare processes will be discussed in a forthcoming paper.

\section{Conclusions}

We have studied the mediation of supersymmetry breaking in unified models
with GUT-sized extra dimensions. The MSSM soft terms receive contributions
both from gravity mediation and from gauge mediation, since such 
models often contain suitable messenger fields with GUT-scale masses. These 
GUT-scale messengers tend to come in large numbers, and they do not have to 
form complete GUT multiplets. The presence of incomplete multiplets turns out
to be crucial in achieving realistic electroweak symmetry breaking. Gauge 
mediation dominates, and as a consequence the superpartner spectrum is quite 
peculiar: The higgsinos can be as light as $100-200$ GeV, because the
$\mu$ term is induced by gravity mediation, while the remaining superpartners
and the heavy Higgs bosons are much heavier. We have presented some sample
spectra, and discussed the implications for naturalness and cosmology. As the
lightest superparticle, the gravitino is a natural dark matter candidate. The
relic abundance of the higgsino NLSP is substantially reduced by coannihilation,
but there is still some tension remaining with the current bounds from primordial 
nucleosynthesis. A study of the resulting collider phenomenology is currently 
in preparation.

\section*{Acknowledgements}
The authors thank L.~Covi, J.~Hasenkamp, S.~Kraml, and R.~Valandro for useful
discussions.

\end{document}